\documentclass[12pt]{article}
\usepackage{ctex}
\usepackage{amssymb,amsmath,color,amsmath,bm}
\usepackage[colorlinks,linkcolor=blue]{hyperref}
\usepackage{geometry}
\usepackage{tabularx}
\usepackage{bm}
\usepackage{authblk}
\usepackage[square, comma, sort&compress, numbers]{natbib}

	\title{\bf Infinite families  $2$-designs from  binary  projective three-weight codes}		
	
	\author{\small Canze Zhu$^{1}$,~~~~Qunying Liao$^2$,
		\thanks{Corresponding author.\\
	\quad Canze Zhu:~canzezhu@163.com\\
	\quad Qunying Liao:~qunyingliao@sicnu.edu.cn\\
	\quad Haibo Liu:~liuhaibo@cuit.edu.cn\\			
		{~~The work of Q.Y. Liao is supported by National Natural Science Foundation of China (No. 12071321)}\\{~~The work of H.B. Liu is supported by the Natural Science Foundation of China (No. 11901062)}}~~~~ ~~~~Haibo Liu$^3$}
	
	\affil[]{\small}
	\date{}
	\newtheorem{theorem}{Theorem}[section]
	
	\newtheorem{lemma}{Lemma}[section]

	\newtheorem{remark}{Remark}[section]

	\catcode`@=11 \@addtoreset{equation}{section} \catcode`@=12
\begin{document}
	\maketitle	
	{\footnotesize
		\centerline{$^1$School of Computer Science and Technology, Dongguan University of Technology, Dongguan, China, 523000}
	} 
	\medskip
	{\footnotesize
		\centerline{$^2$College of Mathematical Science, Sichuan Normal University, Chengdu, China, 610066}
	}	\medskip
{\footnotesize
\centerline{$^3$School of Applied Mathematics, Chengdu University of Information Technology, Chengdu, Sichuan, China, 610225}
}\bigskip
	{\bf Abstract.}
	{\small Combinatorial designs are closely related to linear codes. In recent year, there are a lot of  $t$-designs constructed from certain linear codes. In this paper, we aim to construct  $2$-designs from binary three-weight codes. For any binary  three-weight code $\mathcal{C}$ with length $n$, let $A_{n}(\mathcal{C})$ be the number of codewords in $\mathcal{C}$ with Hamming weight $n$, then we show that $\mathcal{C}$ holds $2$-designs when $\mathcal{C}$ is projective and $A_{n}(\mathcal{C})=1$. Furthermore, by extending some certain binary projective two-weight codes and basing on the defining set method, we construct two classes of binary  projective three-weight codes which are suitable for holding $2$-designs. }\\
	
	\section{Introduction}
	Let $\mathcal{P}$ be a set of $v $ elements, and let $\mathcal{B}$ be a set
	of $r$-subsets of $\mathcal{P}$, where $r,v\in\mathbb{N}^{+}$ with
	$1 \le r \le v.$ 
	The elements of $\mathcal{P}$ are called points, and those  of $\mathcal{B}$ are referred to as blocks.  For $t\in\mathbb{N}^{+}$ with $t \le r$, the incidence structure $\mathbb{D}= (\mathcal{P},\mathcal{B})$, where the incidence relation
	is the set membership, is called a $t$-$(v, r, \lambda)$ design if every $t$-subset of $\mathcal{P}$ is contained in exactly $\lambda$
	elements of $\mathcal{B}$.   
	We usually use $b$ to denote the number of blocks in $\mathcal{B}$. Let $\binom{\mathcal{P}}{r}$ denote the set of all $r$-subsets of $\mathcal{P}$. 
	A $t$-design is called simple if $\mathcal{B}$ does not contain any repeated blocks. In this paper, we consider
	only simple $t$-designs with $v>r>t$. A $t$-$(v, r, \lambda)$ design
	is referred to as a Steiner system if $t \ge 2$ and $\lambda = 1$, and is
	denoted by $S(t, r, v)$. Furthermore, for the $t$-$(v, r, \lambda)$ design, the following equality holds \cite{WV}:
	\begin{align}\label{TD}
	b\binom{r}{t}=\lambda \binom{v}{t}.
	\end{align}
	
	 Let $\mathbb{F}_{q}$ be the finite field with $q$ elements, where $q$ is a prime power. A linear code $\mathcal{C}$ over $\mathbb{F}_q$ with parameters $[n,k,d]$ is a $k$-dimensional subspace of $\mathbb{F}_q^n$ with minimum (Hamming) distance $d$ and length $n$. The dual code of $\mathcal{C}$ is defined as
	 $\mathcal{C}^{\perp}=\{\boldsymbol{c}^{'}\in\mathbb{F}_q^n~|~\langle \boldsymbol{c}^{'},\boldsymbol{c}\rangle=0~\text{for all}~\boldsymbol{c}\in\mathcal{C}\},$ where $\langle\boldsymbol{c}^{'},\boldsymbol{c}\rangle$ is the Euclidean inner product of $\boldsymbol{c}$ and $\boldsymbol{c}^{'}$. It is easy to see that the dimension of $\mathcal{C}^{\perp}$ is $n-k$. $\mathcal{C}$ is said to be projective if the minimum distance of $\mathcal{C}^{\perp}$  is greater than or equal to $3$.  Let $A_i(\mathcal{C})\ (i=0,1,\ldots,n)$ be the number of codewords with Hamming weight $i$ in $\mathcal{C}$.  Then the sequence $(1,A_1(\mathcal{C}),\ldots,A_{n}(\mathcal{C}))$ is called the weight distribution of $\mathcal{C}$, and  the weight enumerator  of $\mathcal{C}$ is defined by the polynomial 
	 \begin{align*}
	 P_\mathcal{C}(z)=1+A_1(\mathcal{C})z+\cdots+A_n(\mathcal{C})z^n.
	 \end{align*}
	 $\mathcal{C}$ is said to be an $l$-weight code if the number of nonzero $A_i(\mathcal{C})$ in the sequence $(A_1(\mathcal{C}),\ldots,A_{n}(\mathcal{C}))$ is equal to $l$.
	 Especially, projective two-weight codes and projective three-weight codes are very interesting as they have close connections with finite projective spaces,  strongly regular graphs, strongly walk regular graphs, s-sum sets,  and so on\cite{RC1986,DT1,P1972,SK,SS,ZZ,ZZ1,ZZ2,ZH2021,ZH2016,ZH20161}.


	There are some close relationships between linear codes and $t$-designs. In particularly, $t$-designs can be constructed from  linear codes satisfying some certain conditions. This construction is introduced below. 
	
	For an $[n,k,d]$ linear code $\mathcal{C}$ over $\mathbb{F}_{q}$,  let the coordinates of a codeword in $\mathcal{C}$ be indexed by $(1,\ldots,n)$. The support of a codeword $\boldsymbol{c}=(c_1,\ldots,c_n)\in \mathcal{C}$ is defined as \begin{align*}
	 \mathrm{Supp}(\boldsymbol{c})=\{1\le i\le n~|~c_i\neq 0\}.
	 \end{align*}
	 For each $r$ with $A_{r}(\mathcal{C})\neq 0$, let $\mathcal{B}_r$ denote the set of supports of all the codewords of Hamming weight $r$ in $\mathcal{C}$. Let $\mathcal{P}=\{1,\ldots,n\}$. The pair $(\mathcal{P},\mathcal{B}_r)$ may be a $t$-$(n,r,\lambda)$ design for some positive integer $t$ and $\lambda$, which is called a support design of $\mathcal{C}$. In other words, we say that the codewords with weight $r$ in $\mathcal{C}$ hold a $t$-$(n,r,\lambda)$ design. When the pair $(\mathcal{P},\mathcal{B}_r)$ holds a simple $t$-$(n,r,\lambda)$ design, there are the following relation \cite{DT2}:
	 \begin{align*}
	 \left|\mathcal{B}_r\right|=\frac{1}{q-1}A_{r}(\mathcal{C}),~~~\lambda\binom{n}{t}=\frac{1}{q-1}A_{r}(\mathcal{C})\binom{r}{t}.
	 \end{align*}
	 Especially, if $\mathcal{C}$ is a binary code, i.e., $q=2$, then
	 \begin{align*}
	 	 \left|\mathcal{B}_r\right|=A_{r}(\mathcal{C}),~~~\lambda\binom{n}{t}=A_{r}(\mathcal{C})\binom{r}{t}.
	 \end{align*}
	 If the weight distributions of $\mathcal{C}$ and $\mathcal{C}^{\perp}$ are both of a particular form, a powerful result due to Assmus and Mattson guarantees that $(\mathcal{P},\mathcal{B}_r)$ is a $t$-design \cite{WV}. Based on Assmus-Mattson, infinite families $t$-designs have been constructed from  certain linear codes \cite{DT1,DT2,DT3,DT4,DT5,DT6,DT7, DT8,DT9}

	The objective of this paper is to construct infinite families $2$-designs from  three-weight linear codes.
	Let $\mathcal{C}$ be a binary projective  three-weight code with length $n$. If $A_{n}(\mathcal{C})>0$, then according to the Assmus-Mattson Theorem, we show that
	 $2$-designs can be obtained from $\mathcal{C}$ and $\mathcal{C}^{\perp}$. Furthermore, we give two constructions of binary three-weight codes satisfying these conditions, one of the constructions is by extending the binary projective two-weight codes, and the other is based on the defining set method.

The remainder of this paper is arranged as follows. In Section $2$, we introduce the  MacWilliam Identity, the Assmus-Mattson Theorem, and some related results of Weil sums. In Section $3$, we show that  $2$-designs can be constructed from  binary  three-weight codes satisfying some conditions. In Section $4$,
by extending some known binary two-weight codes and the defining set method, we construct several classes of binary three-weight codes, which are suitable for holding $2$-designs.
 In section 5, we conclude the whole paper.  
\section{Preliminaries}

In this section, the  MacWilliam Identity, the Assmus-Mattson Theorem, and some related results of Weil sums are presented.

\subsection{The MacWilliam Identity and the Assmus-Mattson Theorem for binary linear codes}
The  MacWilliam Identity and the first two Pless power moments for binary linear codes are given in the following lemmas, respectively, which are needed to calculate  weight distributions for dual codes of binary linear codes.

\begin{lemma}[\cite{WV}, The MacWilliam Identity]\label{PPP}
	Let $\mathcal{C}$ be a binary $[n,k,d]$ code with weight enumerator 
	$$P_\mathcal{C}(z)=1+A_1(\mathcal{C})z+\cdots+A_n(\mathcal{C})z^n,$$ then
	\begin{align*}
	P_\mathcal{C^{\perp}}(z)=2^{-k}(1+z)^{n}P_\mathcal{C}\left(\frac{1-z}{1+z}\right).
	\end{align*}
\end{lemma}

Pless power moments are the equivalence transformation of the  MacWilliam Identity, the first three Pless power moments are listed in the following lemma.
\begin{lemma}[\cite{WV}, The first three Pless power moments]\label{PP}
	For any binary code $\mathcal{C}$ with parameters $[n, k, d]$,  the first three Pless power
	moments are
	\begin{align*}
	&\sum_{j=0}^{n}A_j(\mathcal{C})=2^{k},\\
	&\sum_{j=0}^{n}jA_j(\mathcal{C})=2^{k-1}(n- A_1(\mathcal{C}^{\perp})),
	\end{align*}
	and\begin{align*}
	&\sum_{j=0}^{n}j^2A_j(\mathcal{C})=2^{k-2}\Big(n(n+1)-2nA_1(\mathcal{C}^{\perp})+2A_2(\mathcal{C}^{\perp})\Big).
	\end{align*}
\end{lemma}

The Assmus-Mattson Theorem for binary linear codes is stated as follows, which will be used for the construction of $2$-designs from binary  projective three-weight codes satisfying some conditions.

\begin{lemma}[\cite{WV}, The Assmus-Mattson Theorem]\label{AM} Let $\mathcal{C}$ be a binary $[n,k,d]$ code. Suppose $\mathcal{C}^{\perp}$ has minimum distance $d^{\perp}$. For a fixed integer $t$ with $t<d$, and let $s$ be the number of $i$ with $A_{i}(\mathcal{C}^{\perp})\neq 0$ for $0<i\le n-t$. Suppose $s\le d-t$. Then:\\
	$(1)$	the codewords of weight $i$ in $\mathcal{C}$ hold a $t$-design provided $A_{i}(\mathcal{C})\neq 0$ and $d\le i\le n$, and\\
	$(2)$ the codewords of weight $i$ in $\mathcal{C}^{\perp}$ hold a $t$-design provided $A_{i}(\mathcal{C}^{\perp})\neq 0$ and $d^{\perp}\le i\le n-t$.
\end{lemma}

\subsection{Some known results of Weil sum}
In the following, we list some basic knowledge of additive characters and results of Weil sums, which shall be useful for determining the weight distribution of three-weight codes constructed from the defining set method in Section $4.2$.

Let $q=2^m$, the additive character of $\mathbb{F}_{q}$ is a function from $\mathbb{F}_{q}$ to the multiplicative group $U=\{u\ |\ |u|=1,\ u\in\mathbb{C}\}$, such that $\chi(x+y)=\chi(x)\chi(y)$ for all $x,\ y\in\ \mathbb{F}_{q}$. For each $b\in \mathbb{F}_{q}$, 
\begin{center}
	$\chi_b(x)=(-1)^{\mathrm{Tr}(bx)}$ \quad $(x\in \mathbb{F}_{q})$
\end{center}
defines an additive character of $\mathbb{F}_{q}$,  where $\mathrm{Tr}$ is the absolute trace function from $\mathbb{F}_q$ to $\mathbb{F}_2$. Especially, 
$\chi:=\chi_1$ is called the canonical additive character of $\mathbb{F}_{q}$, and  every additive character of $\mathbb{F}_{q}$ can be written as $\chi_b(x)=\chi(bx)$. 
Furthermore, the orthogonal property for the additive character is given by
\begin{align*}
\sum_{x\in \mathbb{F}_{q}}(-1)^{\mathrm{Tr}(bx)}=\begin{cases}
q,\quad& \text{if}~b=0;\\
0,\quad &\text{otherwise}.
\end{cases}
\end{align*} 

 Weil sums are defined by $\sum\limits_{x\in\mathbb{F}_{q}}\chi(f(x))$ for any $f(x)\in\mathbb{F}_{q}[x]$. For any $u\in\mathbb{N}^{+}$ and $(a,b)\in\mathbb{F}_q^2$,  Coulter  determined the value of Weil sum as
\begin{align*}
S_{u}(a,b)=\sum_{x\in\mathbb{F}_{q}}(-1)^{\mathrm{Tr}(ax^{2^u+1}+bx)}.
\end{align*} 
Before presenting the explicit value of $S_{u}(a,b)$, we fixed some notations as follows:\\

$\bullet$ $e$ is the greatest common divisor of $m$ and $u$, i.e., $e=\gcd (m,u)$.

$\bullet$ $\mathrm{Tr}_e$ is the  trace function from $\mathbb{F}_q$ to $\mathbb{F}_{2^e}$.\\
 
 When $\frac{m}{e}$ is odd, the explicit value of $S_{u}(a,b)$ is given in the following lemmas, which shall be used for calculating the  weight distributions for  binary three-weight linear codes constructed in Section $4.2$.
\begin{lemma}[\cite{RC}, Theorem $4.1$]\label{Ws1}
	If $\frac{m}{e}$ is odd and $a\in\mathbb{F}_{q}^{*}$, then
	\begin{align*}
	S_{u}(a,0)=0.
	\end{align*}
\end{lemma}
\begin{lemma}[\cite{RC}, Theorem $4.2$]\label{Ws2}
If $\frac{m}{e}$ is odd and $(a,b)\in(\mathbb{F}_{q}^{*})^2$, then
\begin{align*}
S_{u}(a,b)=S_{u}(1,b\gamma^{-1}),
\end{align*}
where $\gamma\in\mathbb{F}_q^{*}$ is the unique element satisfying $\gamma^{2^{u+1}}=a$. Furthermore, we have
\begin{align*}
S_{u}(1,b)=\begin{cases}
0,&~\text{if}~\mathrm{Tr}_{e}(b)\neq 1;\\
\pm 2^{\frac{m+e}{2}},&~\text{if}~\mathrm{Tr}_{e}(b)=1.
\end{cases}
\end{align*}
\end{lemma}
\begin{lemma}[\cite{RC}, Theorem $4.6$]\label{Ws20}
If $\frac{m}{e}$ is odd, then
\begin{align*}
S_{u}(1,1)=\left(\frac{~2~}{\frac{m}{e}}\right)^e2^{\frac{m+e}{2}},
\end{align*}
where $\left(\frac{~\cdot~}{\cdot}\right)$ is the Jacobi symbol.
\end{lemma}	
\begin{remark}It is well-known that $\left(\frac{~2~}{s}\right)=(-1)^{\frac{s^2-1}{8}}$, where $s>1$ is odd. Then 	
	\begin{align*}
	S_{u}(1,1)=(-1)^{\frac{m^2-e^2}{8e}}2^{\frac{m+e}{2}}.
	\end{align*}
\end{remark}


\section{$2$-designs from binary three-weight linear codes with special parameters}
In the following lemma,  we determined the weight distributions for a class of binary three-weight codes and its dual codes.
\begin{theorem}\label{TW}
	Let $\mathcal{C}$ be an $[n,k,d]$ binary  projective three-weight code, if $A_{n}(\mathcal{C})>0$, then the three nonzero weights of $\mathcal{C}$ are 
	\begin{align*}
		w_1=\frac{1}{2}\left(n-\sqrt{n\left(\frac{2^{k-1}-n}{2^{k-1}-1}\right)}\right),~w_2=\frac{1}{2}\left(n+\sqrt{n\left(\frac{2^{k-1}-n}{2^{k-1}-1}\right)}\right),~~~w_3=n.
	\end{align*}
	Furthermore, 
	\begin{align*}
	P_\mathcal{C}(z)=1+(2^{k-1}-1)z^{d}+(2^{k-1}-1)z^{n-d}+z^n,
	\end{align*}
	and 
	\begin{align*}
			P_{\mathcal{C}^{\perp}}(z)=1+\sum_{r=2}^{\lfloor \frac{n}{2} \rfloor}A_{2r}(\mathcal{C}^{\perp})z^{2r},
	\end{align*}
	where for any $r=2,\ldots,\left\lfloor \frac{n}{2}\right\rfloor$,
	\begin{align*}
	A_{2r}(\mathcal{C}^{\perp})=2^{-(k-1)}\left(\binom{n}{2r}+(2^{k-1}-1)\sum_{\substack{0\le i\le d\\0\le j\le \lfloor\frac{n}{2}\rfloor-d\\i+j=r}}(-1)^{i}\binom{d}{i}\binom{n-2d}{2j}\right).
	\end{align*}
\end{theorem}

{\bf Proof.} Let $w_1<w_2<w_3$ be three nonzero weights of $\mathcal{C}$, by the assumptions that $\mathcal{C}$  is a binary code and $A_n>0$, we have 
\begin{align*}
	w_1=d,~~w_2=n-d,~~w_3=n~~\text{and}~~A_n=1.
\end{align*}
 Since $\mathcal{C}$ is projective, one has $A_{1}(\mathcal{C}^{\perp})=A_{2}(\mathcal{C}^{\perp})=0$, then by Lemma \ref{PP}, we immediately get $$A_{w_1}(\mathcal{C})=A_{w_2}(\mathcal{C})=2^{k-1}-1.$$ Thus 
\begin{align*}
P_\mathcal{C}(z)=1+(2^{k-1}-1)z^{d}+(2^{k-1}-1)z^{n-d}+z^n.
\end{align*}
Then it follows from Lemma \ref{PPP} that 
\begin{align*}
P_{\mathcal{C}^{\perp}}(z)=&2^{-k}(1+z)^{n}\left(1+(2^{k-1}-1)\left(\frac{1-z}{1+z}\right)^{d}+(2^{k-1}-1)\left(\frac{1-z}{1+z}\right)^{n-d}+\left(\frac{1-z}{1+z}\right)^{n}\right)\\
						=&2^{-k}\left((1+z)^{n}+(1-z)^{n}+(2^{k-1}-1)\Big((1+z)^{n-d}(1-z)^{d}+(1+z)^d(1-z)^{n-d}\Big)\right)\\
						=&2^{-k}\left((1+z)^{n}+(1-z)^{n}+(2^{k-1}-1)\big(1-z^2\big)^d\Big((1+z)^{n-2d}+(1-z)^{n-2d}\Big)\right)\\
						=&2^{-k}\left(2\sum_{r=0}^{\lfloor\frac{n}{2}\rfloor}\binom{n}{2r}z^{2r}+(2^{k-1}-1)\left(\sum_{i=0}^{d}(-1)^i\binom{d}{i}z^{2i}\right)\left(2\sum_{j=0}^{\lfloor\frac{n-2d}{2}\rfloor}\binom{n-2d}{2j}z^{2j}\right)\right)\\
						=&2^{-(k-1)}\left(\sum_{r=0}^{\lfloor\frac{n}{2}\rfloor}\binom{n}{2r}z^{2r}+(2^{k-1}-1)\sum_{\substack{0\le i\le d\\0\le j\le \lfloor\frac{n}{2}\rfloor-d\\i+j=r}}(-1)^{i}\binom{d}{i}\binom{n-2d}{2j}z^{2r}\right)\\
						=&1+\sum_{r=1}^{\lfloor \frac{n}{2} \rfloor}A_{2r}(\mathcal{C}^{\perp})z^{2r},
\end{align*}
where for any $r=2,\ldots,\left\lfloor \frac{n}{2}\right\rfloor$,
\begin{align*}
&A_{2r}(\mathcal{C}^{\perp})=2^{-(k-1)}\Bigg(\binom{n}{2r}+(2^{k-1}-1)\sum_{\substack{0\le i\le d\\0\le j\le \lfloor\frac{n}{2}\rfloor-d\\i+j=r}}(-1)^{i}\binom{d}{i}\binom{n-2d}{2j}\Bigg).
\end{align*}
Now by $A_2(\mathcal{C}^{\perp})=0$, one has
\begin{align*}
2^{-(k-1)}\Bigg(\binom{n}{2}+(2^{k-1}-1)\sum_{\substack{0\le i\le d\\0\le j\le \lfloor\frac{n}{2}\rfloor-d\\i+j=1}}(-1)^{i}\binom{d}{i}\binom{n-2d}{2j}\Bigg)=0,
\end{align*}
i.e.,
\begin{align*}
	4d^2-4nd+\frac{2^{k-1}}{2^{k-1}-1}n(n-1)=0.
\end{align*}
Thus 
\begin{align*}
	w_1=d=\frac{1}{2}\left(n-\sqrt{n\left(\frac{2^{k-1}-n}{2^{k-1}-1}\right)}\right)~~\text{and}~~w_2=n-d=\frac{1}{2}\left(n+\sqrt{n\left(\frac{2^{k-1}-n}{2^{k-1}-1}\right)}\right).
\end{align*}
So far, we get the desired results. $\hfill\Box$\\

In the following theorem, we give the connection between binary  projective three-weight codes and $2$-designs.
\begin{theorem}\label{TWD}
	
	 If  $\mathcal{C}$ is a binary  projective three-weight code with parameters $[n,k,d]$  and $A_{n}(\mathcal{C})=1$, then the following two assertions holds.

	$(1)$ $\mathcal{C}$ has three nonzero codewords $$w_1=\frac{1}{2}\left(n-\sqrt{n\left(\frac{2^{k-1}-n}{2^{k-1}-1}\right)}\right),~w_2=\frac{1}{2}\left(n+\sqrt{n\left(\frac{2^{k-1}-n}{2^{k-1}-1}\right)}\right),~w_3=n,$$ and the codewords in $\mathcal{C}$ with weight $w_i$ $(i=1,2)$ hold a $2$-$(n,w_i,\lambda_{w_i})$ $(i=1,2)$ design, where
	$$\lambda_{w_i}=\frac{(2^{k-1}-1)w_i(w_i-1)}{n(n-1)}~~~~(i=1,2).$$
	
	$(2)$ For $r=2,\ldots, \lfloor\frac{n}{2}\rfloor-1$, the codewords  in $\mathcal{C}^{\perp}$ with weight $2r$ hold a $2$-$(n,2r,\lambda_r^{\perp})$ design  provided $A_{2r}(\mathcal{C}^{\perp})\neq 0$, where
	\begin{align*}
	A_{2r}(\mathcal{C}^{\perp})=2^{-(k-1)}\Bigg(\binom{n}{2r}+(2^{k-1}-1)\sum_{\substack{0\le i\le d\\0\le j\le \lfloor\frac{n}{2}\rfloor-d\\i+j=r}}(-1)^{i}\binom{d}{i}\binom{n-2d}{2j}\Bigg)
	\end{align*}and
	\begin{align*}
		\lambda_r^{\perp}=\frac{2r(2r-1)}{n(n-1)}\cdot A_{2r}(\mathcal{C}^{\perp}).
	\end{align*}
	
\end{theorem}

{\bf Proof.}  We can check that $\mathcal{C}^{\perp}$ is suitable for conditions in  Lemma \ref{AM} when $t=2$, thus the corresponding $2$-$(n,2r,\lambda^{\perp}_r)$ $(r=2,\ldots,\lfloor\frac{n}{2}\rfloor-1)$ design and $2$-$(n,w_i,\lambda_{w_i})$ $(i=1,2)$ design  are given, respectively. Furthermore, $\lambda^{\perp}_r$ and $\lambda_{w_i}$ are determined by $(\ref{TD})$ and Theorem $\ref{TW}$ directly. $\hfill\Box$\\

Theorem \ref{TWD} implies that  binary three-weight linear codes with special parameters can hold $2$-designs. However, there are a lot of results about constructions of binary three-weight linear codes \cite{DD2014,HY2015,QTH2015,LCXM2018,LYWY2019,CZZX2020,HLZ2020,ZZ1,YY2021}, we have not found any binary three-weight codes that can satisfy the conditions in Theorem \ref{TWD}. In the next section, we shall construct two classes of binary three-weight codes that satisfy the conditions in Theorem \ref{TWD}.

\section{Constructions for  binary  projective three-weight codes}
In this section, we give two constructions for binary three-weight codes that are suitable for the conditions in Theorem \ref{TWD}, one of the constructions  is by extending the binary  projective 
two-weight codes, and the other is based on the defining set method.
\subsection{Three-weight codes from some known two-weight codes}
In this subsection, for any binary linear code $\mathcal{C}$ with length $n$ and dimension $k$, set $\overline{\boldsymbol{c}}=(c_1,\ldots,c_{n},0)$ for any $\boldsymbol{c}=(c_1,\ldots,c_n)\in\mathcal{C}$. Then the extended code of $\mathcal{C}$ is defined as
\begin{align*}
\overline{\mathcal{C}}=\{\overline{\boldsymbol{c}}~|~\boldsymbol{c}\in\mathcal{C}\}\cup\{\boldsymbol{1}+\overline{\boldsymbol{c}}~|~\boldsymbol{c}\in\mathcal{C}\},
\end{align*}
where  $\boldsymbol{1}=(1,\ldots,1)\in\mathbb{F}_{2}^{n+1}$. 
It is easy to check that $\overline{\mathcal{C}}$ is a linear code with length $n+1$ and dimension $k+1$. 

In the following theorem, let $\mathcal{C}$ be a two-weight code satisfying some conditions, we show that $\overline{\mathcal{C}}$ is a three-weight code which is suitable for the conditions in Theorem \ref{TWD}. 
\begin{theorem}\label{ETW}
For $n>2$,	let $\mathcal{C}$ be an $[n-1,k-1,d]$ binary projective two-weight code with two nonzero weights $w_1$ and $w_2$ satisfying $w_1=d$ and $w_2=n-d$, then $\overline{\mathcal{C}}$ is an $[n,k,d]$ binary  projective three-weight code with $A_{n}(\overline{\mathcal{C}})=1$. 
\end{theorem}

{\bf Proof}. For any $\boldsymbol{c}\in\mathcal{C}$, let $wt_H(\boldsymbol{c})$ (resp., $wt_H(\overline{\boldsymbol{c}})$) be the Hamming weight of $\boldsymbol{c}$ (resp., $wt_H(\overline{\boldsymbol{c}})$). 
By the assumption, we know that $\mathcal{C}$ is a binary two-weight code with two nonzero weights $w_1$ and $w_2$ satisfying $w_1+w_2=n$. Notice that $wt_H(\overline{\boldsymbol{c}})=wt_H(\boldsymbol{c})=w$, then
\begin{align*}
wt_H(\boldsymbol{1}+\overline{\boldsymbol{c}})=n-wt_H(\overline{\boldsymbol{c}})
=\begin{cases}
	n,&\text{if~}wt_H({\boldsymbol{c}})=0,\\
	n-d,&\text{if~}wt_H({\boldsymbol{c}})=w_1;\\
	d,&\text{if~}wt_H({\boldsymbol{c}})=w_2.
	\end{cases}
\end{align*}
Hence $\overline{\mathcal{C}}$ is an $[n,k,d]$ binary three-weight code with three nonzero weights $w_1=d$, $w_2=n-d$ and $w_3=n$. Furthermore,  $\overline{\mathcal{C}}$ is projective follows from the fact that $\mathcal{C}$ is  projective. $\hfill\Box$\\

By searching the two-weight codes constructed in \cite{HY2015,QTH2015,LCXM2018,LYWY2019,CZZX2020,HLZ2020}, we find two classes of two-weight codes that satisfy the conditions in Theorem \ref{ETW}, which are presented in the following two lemmas.
\begin{lemma}[ \cite{LCXM2018}, Theorem $1(ii)$,~~\cite{QTH2015}, Theorem $2.1$]\label{TW1}
	Let $k\ge 2$ be an integer, then there exists a binary projective two-weight code with parameters $[2^{2k-1}-2^{k-1}-1,2k,2^{2k-2}-2^{k-1}]$ and two nonzero weights $w_1=2^{2k-2}-2^{k-1}$ and $w_2=2^{2k-2}$. 
\end{lemma}
\begin{lemma}[\cite{LCXM2018}, Theorem $3(ii)$]\label{TW2}
	Let $k\ge 3$ be odd, then there exists a binary projective two-weight code with parameters $[2^{2k-1}+2^{k-1}-1,2k,2^{2k-2}]$ and two nonzero weights $w_1=2^{2k-2}$ and $w_2=2^{2k-2}+2^{k-1}$. 
\end{lemma}

By Theorem $\ref{ETW}$ and Lemmas \ref{TW1}-\ref{TW2}, we immediately have the following two theorems.
\begin{theorem}\label{t31}
		Let $m\ge 5$ be odd, then there exists a binary  projective three-weight code $\mathcal{C}_1$ with parameters $[2^{m-2}-2^{\frac{m-3}{2}},m,2^{m-3}-2^{\frac{m-3}{2}}]$ and $A_{2^{m-2}-2^{\frac{m-3}{2}}}(\mathcal{C}_1)=1$.
\end{theorem}
\begin{theorem}\label{t32}
	Let $m\ge 7$ be an integer with $m\equiv 3~(\mathrm{mod}~4)$, then there exists a binary  projective three-weight code $\mathcal{C}_2$ with parameters $[2^{m-2}+2^{\frac{m-3}{2}},m,2^{m-3}]$ and $A_{2^{m-2}+2^{\frac{m-3}{2}}}(\mathcal{C}_2)=1$.
\end{theorem}

In the next subsection, by the defining set method, we construct two classes of binary three-weight codes with parameters the same as those of the codes presented in Theorems \ref{t31}-\ref{t32}, and based on these constructions, the restriction of $m\equiv 3~(\mathrm{mod}~4)$ in Theorem \ref{t32} can be omitted.

\subsection{Two classes of binary three-weight codes from defining sets }
In 2015, Ding et al. proposed the defining set construction method to construct few-weight linear codes \cite{DD2014}. Let $D=\{ d_1,\ldots, d_n \}\subseteq\mathbb{F}_{p^{m}}^*$, the linear code over $\mathbb{F}_{p}$ is defined as
\begin{align*}
\mathcal{C}_{D}=\big\{\mathbf{c}(b)=\big(\mathrm{Tr}(bd_1),\ldots,\mathrm{Tr}(bd_n)\big)~\big{|}~ b\in\mathbb{F}_{p^{m}}\big\}.
\end{align*}
Here  $D$ is called the defining set of $\mathcal{C}_{D}$.

In this subsection,  based on the defining set method, we construct two classes of binary  projective three-weight linear codes, which are suitable to get $2$-designs.
\subsubsection{Main results}
 Let $q=2^m$ with $m\in\mathbb{N}^{+}$. For any $u\in \mathbb{N}^{+}$ and $\rho\in\mathbb{F}_2$, set
\begin{align*}
	D_{\rho}=\left\{x\in \mathbb{F}_{q}~|~\mathrm{Tr}(x)=1,~\mathrm{Tr}(x^{2^u+1})=\rho\right\}.
\end{align*}
In the following theorems, we show that $\mathcal{C}_{D_{\rho}}$ is a  projective three-weight linear code and determine the weight distribution of $\mathcal{C}_{D_{\rho}}$. 
\begin{theorem}\label{t1}
For any odd $m\ge 5$, let $u\in N^{+}$ with $\gcd(u,m)=1$. If $\frac{m^2-1}{8}$ is even and $\rho=0$ or  $\frac{m^2-1}{8}$ is odd and $\rho=1$, 
 then $\mathcal{C}_{D_{\rho}}$ is a binary  projective three-weight code with parameters $\left[2^{m-2}-2^{\frac{m-3}{2}},m,2^{m-3}-2^{\frac{m-3}{2}}\right]$ and the weight enumerator
\begin{align*}\mathcal{P}_{\mathcal{C}_{D_{\rho}}}(z)=
1+\big(2^{m-1}-1\big)z^{2^{m-3}-2^{\frac{m-3}{2}}}+\big(2^{m-1}-1\big)z^{2^{m-3}}+z^{2^{m-2}-2^{\frac{m-3}{2}}}.
\end{align*}
\end{theorem}\begin{theorem}\label{t2}
For any odd $m\ge 5$, let $u\in N^{+}$ with $\gcd(u,m)=1$. If $\frac{m^2-1}{8}$ is even and $\rho=1$ or  $\frac{m^2-1}{8}$ is odd and $\rho=0$, 
then $\mathcal{C}_{D_{\rho}}$ is a binary  projective three-weight code with parameters $\left[2^{m-2}+2^{\frac{m-3}{2}},m,2^{m-3}\right]$ and the weight enumerator
\begin{align*}\mathcal{P}_{\mathcal{C}_{D_{\rho}}}(z)=
1+\big(2^{m-1}-1\big)z^{2^{m-3}}+\big(2^{m-1}-1\big)z^{2^{m-3}+2^{\frac{m-3}{2}}}+z^{2^{m-2}+2^{\frac{m-3}{2}}}.
\end{align*}
\end{theorem}
\begin{remark}
	The parameters of $\mathcal{C}_{D_{\rho}}$ in Theorem \ref{t1} are the same as those of the code presented in Theorem \ref{t31}.
\end{remark}
\begin{remark}
	By Theorem \ref{t2}, the restriction  $m\equiv 3~(\mathrm{mod}~4)$ for $m$ in Theorem \ref{t32} can be omitted. 
\end{remark}
\begin{remark}
	In Theorems \ref{t1}-\ref{t2}, we focus on  $\mathcal{C}_{D_{\rho}}$ for the case $2\nmid m$ and $\gcd(u,m)=1$, and show that  $\mathcal{C}_{D_{\rho}}$ is a  projective three-weight linear code. For the rest of the case,
	in the similar proofs as those of  Theorems \ref{t1}-\ref{t2}, we can get  $\mathcal{C}_{D_{\rho}}$ is a  four-weight linear code. Hence, we do not consider this case further.
\end{remark}

By Theorem \ref{TWD} and Theorems \ref{t1}-\ref{t2}, we immediately present infinite families $2$-designs in the following two theorems.
\begin{theorem}
	Let $\mathcal{C}_{D_{\rho}}$ be the code given in Theorem \ref{t1}. Then the following two assertions hold.
	
	$(1)$ The codewords  with weight $2^{m-3}-2^{\frac{m-3}{2}}$  in $\mathcal{C}_{D_{\rho}}$ hold a $2$-$(2^{m-2}-2^{\frac{m-3}{2}},2^{m-3}-2^{\frac{m-3}{2}},2^{m-3}-2^{\frac{m-3}{2}}-1)$ design, and the codewords with weight $2^{m-3}$ in $\mathcal{C}_{D_{\rho}}$  hold a $2$-$(2^{m-2}-2^{\frac{m-3}{2}},2^{m-3},2^{\frac{m-3}{2}}(2^{\frac{m-1}{2}}+1)(2^{m-3}-1))$ design.
	
	$(2)$ For any $r=2,3,\ldots,2^{m-3}-2^{\frac{m-5}{2}}$, the codewords  with  weight $2r$ in $\mathcal{C}_{D_{\rho}}^{\perp}$ hold a $2$-$\left(2^{m-2}-2^{\frac{m-3}{2}},2r,\frac{2r(2r-1)\cdot A_{2r}(\mathcal{C}_{D_{\rho}}^{\perp})}{(2^{m-2}-2^{\frac{m-3}{2}})(2^{m-2}-2^{\frac{m-3}{2}}-1)}\right)$ design provided $A_{2r}(\mathcal{C}_{D_{\rho}}^{\perp})>0$, where
	{\small\begin{align*}
	A_{2r}(\mathcal{C}_{D_{\rho}}^{\perp})=2^{-(m-1)}\left(\binom{2^{m-2}-2^{\frac{m-3}{2}}}{2r}+(2^{m-1}-1)\sum_{\substack{0\le i\le 2^{m-3}-2^{\frac{m-3}{2}}\\0\le j\le 2^{\frac{m-5}{2}}\\i+j=r}}(-1)^{i}\binom{2^{m-3}-2^{\frac{m-3}{2}}}{i}\binom{2^{\frac{m-5}{2}}}{2j}\right).
	\end{align*}}
\end{theorem}
\begin{theorem}
	Let $\mathcal{C}_{D_{\rho}}$ be the code given in  Theorem \ref{t2}. Then the following two assertions hold.
		
		$(1)$ The codewords in $\mathcal{C}_{D_{\rho}}$ with weight $2^{m-3}$ holds a $2$-$(2^{m-2}+2^{\frac{m-3}{2}},2^{m-3},2^{m-3}-2^{\frac{m-3}{2}})$ design, and the codewords in $\mathcal{C}_{D_{\rho}}$ with weight $2^{m-3}+2^{\frac{m-3}{2}}$ holds a $2$-$(2^{m-2}+2^{\frac{m-3}{2}},2^{m-3}+2^{\frac{m-3}{2}},2^{m-3}+2^{\frac{m-3}{2}}-1)$ design.
		
		$(2)$ For any $r=2,3,\ldots,2^{m-3}+2^{\frac{m-5}{2}}$, the codewords in $\mathcal{C}_{D_{\rho}}^{\perp}$ with  weight $2r$ hold a $2$-$\left(2^{m-2}+2^{\frac{m-3}{2}},2r,\frac{2r(2r-1)\cdot A_{2r}(\mathcal{C}_{D_{\rho}}^{\perp})}{(2^{m-2}+2^{\frac{m-3}{2}})(2^{m-2}+2^{\frac{m-3}{2}}-1)}\right)$ design provided $A_{2r}(\mathcal{C}_{D_{\rho}}^{\perp})>0$, where
		{\small\begin{align*}
			A_{2r}(\mathcal{C}_{D_{\rho}}^{\perp})=2^{-(m-1)}\left(\binom{2^{m-2}+2^{\frac{m-3}{2}}}{2r}+(2^{m-1}-1)\sum_{\substack{0\le i\le 2^{m-3}\\0\le j\le 2^{\frac{m-5}{2}}\\i+j=r}}(-1)^{i}\binom{2^{m-3}}{i}\binom{2^{\frac{m-5}{2}}}{2j}\right).
			\end{align*}}
\end{theorem}
	
\subsubsection{The proofs of Theorems \ref{t1}-\ref{t2}}
For any codeword $\boldsymbol{c}_{b}\in\mathcal{C}_{D_{\rho}}$, let $D_{\rho}=\{d_1,\ldots,d_n\}$, recall that
\begin{align*}
\boldsymbol{c}_{b}=(\mathrm{Tr}(bd_1),\ldots,\mathrm{Tr}(bd_n)).
\end{align*}
It is easy to see that the length of $\boldsymbol{c}_b$ is  $|D_{\rho}|$, and the Hamming weight of $\boldsymbol{c}_b$ can be given as  
\begin{align}\label{wth}
	wt_H(\boldsymbol{c}_b)=\Big|\{ x \in D_{\rho}~|~\mathrm{Tr}(bx)=1\}\Big|.
\end{align}
Now, we determine the length and the Hamming weight of  $\boldsymbol{c}_b$ in Lemmas \ref{L1}-\ref{L2}, respectively.

\begin{lemma}\label{L1}
	For any odd $m\ge 5$, let $u\in N^{+}$ with $\gcd(u,m)=1$,  then
		\begin{align*}
		|D_{\rho}|
		=&\begin{cases}
		2^{m-2}-(-1)^{\frac{m^2-1}{8}}2^{\frac{m-3}{2}},~&\text{if}~\rho=0;\\
		2^{m-2}+(-1)^{\frac{m^2-1}{8}}2^{\frac{m-3}{2}},~&\text{if}~\rho=1.
		\end{cases}
		\end{align*}
\end{lemma}

{\bf Proof}. It follows from the orthogonal property of additive characters that
\begin{align*}
		|D_{\rho}|=&\sum_{x\in\mathbb{F}_q}\left(\frac{1}{2}\sum_{z_1\in \mathbb{F}_2}(-1)^{z_1\left(\mathrm{Tr}(x)-1\right)}\right)\left(\frac{1}{2}\sum_{z_2\in \mathbb{F}_2}(-1)^{z_2\left(\mathrm{Tr}(x^{2^u+1})-\rho\right)}\right)\\
		=&2^{m-2}+\frac{1}{4}\sum_{x\in\mathbb{F}_q}(-1)^{\mathrm{Tr}(x)-1}+\frac{1}{4}\sum_{x\in\mathbb{F}_q}(-1)^{\mathrm{Tr}(x^{2^{u}+1})-\rho}
		+\frac{1}{4}\sum_{x\in\mathbb{F}_q}(-1)^{\mathrm{Tr}(x^{2^{u}+1}+x)-(1+\rho)}\\
		=&\begin{cases}
		2^{m-2}+\frac{1}{4}S_{u}(1,0)-\frac{1}{4}S_{u}(1,1),~&\text{if}~\rho=0;\\
		2^{m-2}-\frac{1}{4}S_{u}(1,0)+\frac{1}{4}S_{u}(1,1),~&\text{if}~\rho=1.
		\end{cases}
\end{align*}
By Lemmas \ref{Ws1}-\ref{Ws20}, we have $S_{u}(1,0)=0$ and $S_{u}(1,1)=(-1)^{\frac{m^2-v^2}{8v}}2^{\frac{m+v}{2}}$, respectively, where $v=\gcd(m,u)=1$. Thus we get the desired results.
$\hfill\Box$

\begin{lemma}\label{L2}
	For any odd $m\ge 5$, let $u\in N^{+}$ with $\gcd(u,m)=1$. Let $b\in\mathbb{F}_{q}^{*}$, then the following two assertions hold.
	
	$(1)$~If $b=1$, then $wt_H(\boldsymbol{c}_1)=|D_{\rho}|$.
	
	$(2)$~If $b\in \mathbb{F}_{q}\backslash\mathbb{F}_2$, then $wt_H(\boldsymbol{c}_b)\in\left\{\frac{\left|D_{\rho}\right|}{2}-2^{\frac{m-5}{2}},\frac{\left|D_{\rho}\right|}{2}+2^{\frac{m-5}{2}}\right\}$.
\end{lemma}

{\bf Proof}. For $b=1$, by $(\ref{wth})$ and the definition of $D_{\rho}$, we  have $wt_H(\boldsymbol{c}_b)=|D_{\rho}|$.

For $b\in\mathbb{F}_{q}\backslash\mathbb{F}_2$, it follows from $(\ref{wth})$ and the orthogonal property of additive characters that
\begin{align*}
\begin{aligned}
wt_H(\boldsymbol{c}_b)=&\sum_{x\in D_{\rho}}\left(\frac{1}{2}\sum_{y\in \mathbb{F}_2}(-1)^{y\left(\mathrm{Tr}(bx)-1\right)}\right)\\
=&\frac{|D_\rho|}{2}+\frac{1}{2}\sum_{x\in D_{\rho}}\left((-1)^{\mathrm{Tr}(bx)-1}\right)\\
=&\frac{|D_\rho|}{2}+\frac{1}{2}\sum_{x\in \mathbb{F}_{q}}\left((-1)^{\mathrm{Tr}(bx)-1}\right)\left(\frac{1}{2}\sum_{z_1\in \mathbb{F}_2}(-1)^{z_1\left(\mathrm{Tr}(x)-1\right)}\right)\left(\frac{1}{2}\sum_{z_2\in \mathbb{F}_2}(-1)^{z_2\left(\mathrm{Tr}(x^{2^u+1})-\rho\right)}\right)\\
=&\frac{|D_\rho|}{2}-\frac{1}{8}\sum_{x\in \mathbb{F}_{q}}(-1)^{\mathrm{Tr}(bx)}\left(1-(-1)^{\mathrm{Tr}(x)}\right)\left(1+(-1)^{\rho}\cdot(-1)^{\mathrm{Tr}(x^{2^u+1})}\right)\\
=&\frac{|D_\rho|}{2}-\frac{1}{8}\sum_{x\in \mathbb{F}_{q}}(-1)^{\mathrm{Tr}(bx)}+\frac{1}{8}\sum_{x\in \mathbb{F}_{q}}(-1)^{\mathrm{Tr}((b+1)x)}\\
&\qquad-\frac{1}{8}(-1)^{\rho}\cdot\sum_{x\in \mathbb{F}_{q}}(-1)^{\mathrm{Tr}\left(x^{2^u+1}+bx\right)}+\frac{1}{8}(-1)^{\rho}\cdot\sum_{x\in \mathbb{F}_{q}}(-1)^{\mathrm{Tr}\left(x^{2^u+1}+(b+1)x\right)}\\
=&\frac{|D_\rho|}{2}-\frac{1}{8}(-1)^{\rho}\cdot\left(\sum_{x\in \mathbb{F}_{q}}(-1)^{\mathrm{Tr}\left(x^{2^u+1}+bx\right)}-\sum_{x\in \mathbb{F}_{q}}(-1)^{\mathrm{Tr}\left(x^{2^u+1}+(b+1)x\right)}\right).\end{aligned}
\end{align*}
Since $m$ is odd and $e=\gcd(m,u)=1$, we have $\mathrm{Tr}_e(1)=1$. Thus $$\big(\mathrm{Tr}(b), \mathrm{Tr}(b+1)\big)\in\big\{(0,1),(1,0)\big\}.$$
Then by Lemma \ref{Ws2}, we can get
\begin{align*}
	\left(\sum_{x\in \mathbb{F}_{q}}(-1)^{\mathrm{Tr}\left(x^{2^u+1}+bx\right)}-\sum_{x\in \mathbb{F}_{q}}(-1)^{\mathrm{Tr}\left(x^{2^u+1}+(b+1)x\right)}\right)\in\left\{-2^{\frac{m+1}{2}},2^{\frac{m+1}{2}}\right\},
\end{align*}
which leads
\begin{align*}
	wt_H(\boldsymbol{c}_b)\in\left\{\frac{\left|D_{\rho}\right|}{2}-2^{\frac{m-5}{2}},\frac{\left|D_{\rho}\right|}{2}+2^{\frac{m-5}{2}}\right\}.
\end{align*}
$\hfill\Box$\\

In the following, based on Lemma \ref{TW} and Lemmas \ref{L1}-\ref{L2},  the proofs of Theorems \ref{t1}-\ref{t2} are presented.

{\bf The proofs of Theorems \ref{t1}-\ref{t2}:}

By Lemma \ref{L1}, we know that the length of $\mathcal{C}_{D_{\rho}}$ is 
\begin{align}\label{PP1}
\begin{aligned}
|D_{\rho}|=\begin{cases}
2^{m-2}-(-1)^{\frac{m^2-1}{8}}2^{\frac{m-3}{2}},~&\text{if}~\rho=0;\\
2^{m-2}+(-1)^{\frac{m^2-1}{8}}2^{\frac{m-3}{2}},~&\text{if}~\rho=1.\\
\end{cases}
\end{aligned}
\end{align}
Furthermore, it follows from Lemma \ref{L2} that  $\mathcal{C}_{D_{\rho}}$ is a three-weight linear code with parameters $\left[|D_{\rho}|,m,\frac{|D_{\rho}|}{2}-2^{\frac{m-5}{2}}\right]$, and the three nonzero weights are
\begin{align}\label{PP2}
	w_1=\frac{|D_{\rho}|}{2}-2^{\frac{m-5}{2}}, ~~w_2=\frac{|D_{\rho}|}{2}+2^{\frac{m-5}{2}},~~w_3=|D_{\rho}|. 
\end{align}Since $0\notin D_{\rho}$, the minimum distance of $\mathcal{C}_{D_{\rho}}^{\perp}$ cannot be one. Similarly,
since $D_{\rho}$ is not a multiset, i.e.,  two elements $d_i$ and $d_j$ in $D_{\rho}$ must be distinct if $i\neq j$, so the minimum distance of $\mathcal{C}_{D_{\rho}}^{\perp}$ cannot be two. Thus  $\mathcal{C}_{D_{\rho}}$ is projective. Then by Theorem \ref{TW}, we can get
\begin{align}\label{PP3}
		\begin{aligned}	
		\mathcal{P}_{\mathcal{C}_{D_{\rho}}}(z)=\begin{cases}
		1+\big(2^{m-1}-1\big)\Big(z^{2^{m-3}-(I_m+1)2^{\frac{m-5}{2}}}+z^{2^{m-3}-(I_m-1)2^{\frac{m-5}{2}}}\Big)+z^{2^{m-2}-I_m2^{\frac{m-3}{2}}},&\text{if~}\rho=0;\\
		1+\big(2^{m-1}-1\big)\Big(z^{2^{m-3}+(I_m+1)2^{\frac{m-5}{2}}}+z^{2^{m-3}+(I_m-1)2^{\frac{m-5}{2}}}\Big)+z^{2^{m-2}+I_m2^{\frac{m-3}{2}}},&\text{if~}\rho=1,\\
		\end{cases}
		\end{aligned}
\end{align}
where $I_m=(-1)^{\frac{m^2-1}{8}}$.
So far, by $(\ref{PP1})$-$(\ref{PP3})$, we complete the proofs of Theorems \ref{t1}-\ref{t2}.
$\hfill\Box$\\

\section{Conclusions}
In this paper,  the connection between $2$-designs and binary three-weight codes satisfying some conditions are given. Furthermore,  by extending some certain binary projective two-weight codes and based on the defining set method, two classes of binary  projective three-weight codes are constructed, and then infinite families $2$-designs are constructed from these binary three-weight codes.

\noindent{\large\bf Conflict of interest~} The authors have no conflicts of interest to declare that are relevant to the content of this
article.\\

\noindent{\large\bf Data Availability~} Not applicable.\\

\noindent{\large\bf Code Availability~} Not applicable.\\

\noindent{\large\bf Ethical approval~}Not applicable.\\

\noindent{\large\bf Consent to participate~}Not applicable.\\

\noindent{\large\bf Consent for publication~}Not applicable.\\

\end{document}